\documentclass[conference]{IEEEtran}
\IEEEoverridecommandlockouts
\usepackage{cite}
\usepackage{amsmath,amssymb,amsfonts}
\usepackage{algorithmic}
\usepackage{graphicx}
\usepackage{textcomp}
\usepackage{xcolor}
\usepackage{booktabs}

\usepackage{algorithm}
\usepackage{algorithmic}

\usepackage[hyphens]{url}

\def\BibTeX{{\rm B\kern-.05em{\sc i\kern-.025em b}\kern-.08em
    T\kern-.1667em\lower.7ex\hbox{E}\kern-.125emX}}
\begin{document}

\title{One Node at a Time: Node-Level Network Classification}

\author{\IEEEauthorblockN{Saray Shai}
\IEEEauthorblockA{
\textit{Department of Mathematics} \\ \textit{and Computer Science} \\
\textit{Wesleyan University}\\
Middletown, CT USA \\
sshai@wesleyan.edu}
\and
\IEEEauthorblockN{Isaac Jacobs}
\IEEEauthorblockA{
\textit{Department of Mathematics} \\ \textit{and Computer Science} \\
\textit{Wesleyan University}\\
Middletown, CT USA \\
hjacobs@wesleyan.edu}
\and
\IEEEauthorblockN{Peter J. Mucha}
\IEEEauthorblockA{\textit{Department of Mathematics} \\
\textit{Dartmouth College}\\
Hanover, NH USA \\
peter.j.mucha@dartmouth.edu}
}

\maketitle

\begin{abstract}
  Network classification aims to group networks (or graphs) into distinct categories based on their structure. We study the connection between classification of a network and of its constituent nodes, and whether nodes from networks in different groups are distinguishable based on structural node characteristics such as centrality and clustering coefficient. We demonstrate, using various network datasets and random network models, that a classifier can be trained to accurately predict the network category of a given node (without seeing the whole network), implying that complex networks display distinct structural patterns even at the node level. Finally, we discuss two applications of node-level network classification: (i) whole-network classification from small samples of nodes, and (ii) network bootstrapping. 
\end{abstract}


\section{Introduction}
Network (or graph) classification aims to group networks into two or more categories based on their structural properties (and potentially additional metadata)~\cite{cai2018comprehensive}. Such classification tasks rely on ways to measure similarities between structured objects (here, networks), which can then be used to  learn a model that correctly separates networks into distinct classes. 
There are two main approaches to compare the structure of two given networks. 
The more traditional network science approach is based on comparing the statistical properties of interpretable network characteristics such as degree distribution, community structure, and triangle distribution~\cite{baskerville2006subgraph,berlingerio2012netsimile,gallos2014revealing,onnela2012taxonomies,wills2020metrics}.

For example, one can calculate the Kolmogorov-Smirnov statistic between the degree distributions of the two networks, or between more advanced measures such as the distribution of edge density in small subgraphs~\cite{gallos2014revealing}.
The intuition behind this approach is that networks from different categories should have distinct large-scale statistical properties, but knowing which specific properties are ``right'' for the application in hand often requires domain-specific knowledge.

Alternatively, one can embed the networks in a low-dimensional space and calculate the distance between the vectors in the projected space~\cite{yan2006graph}.
For example, kernel methods, which measure pairwise similarity with a kernel function, have become increasingly popular~\cite{kriege2020survey,yanardag2015deep}. In short, graph kernel functions recursively decompose the graph into sub-structures (e.g.\  graphlets~\cite{prvzulj2007biological,shervashidze2009efficient}, subtrees~\cite{shervashidze2009fast,shervashidze2011weisfeiler} or paths~\cite{borgwardt2005shortest,kashima2003marginalized}) and define similarity as the inner product between vectors encoding the frequency of those sub-structures. Other popular methods include factorizing the adjacency matrix~\cite{krzakala2013spectral,lee2001algorithms, newman2006finding, ou2016asymmetric, wang2017community} or encoding it using a multi-layered learning architecture~\cite{defferrard2016convolutional,kipf2016semi,niepert2016learning,scarselli2008graph}. 
The intuition behind these methods is to use advanced tools (such as deep neural networks or kernel functions) to automatically reduce networks into condensed vectors.

Another recently suggested ``hybrid'' approach~\cite{barnett2019endnote} uses a selection of interpretable network features (such as average degree, number of triangles, average clustering coefficient, and degree assortativity) and a classifier which automatically determines feature importance in telling networks apart (such as random forests). In Section~\ref{sec:applications}, we compare our whole-network classification results with this ``feature-based'' classification~\cite{barnett2019endnote} and with Deep Graph Kernels~\cite{yanardag2015deep}.

In this paper, we ask the following question: are nodes from distinct networks (belonging to distinguishable categories) themselves distinguishable? 
{\bf That is, given two networks $\bold{G_1, G_2}$, classified as category $\bold{A}$ and category $\bold{B}$ respectively, can we train a model to distinguish between individual nodes, $\bold{u}$ from $\bold{G_1}$ and $\bold{v}$ from $\bold{G_2}$, based on their structural properties alone?}
For example, given $\langle$degree, clustering coefficient, $\ldots\rangle$ values can a classifier predict that $u$ ($v$) comes from a network in class $A$ ($B$)? 

This question is particularly interesting since one can easily imagine a scenario where (large) networks ``look'' different on a global scale but individual nodes in those networks have similar properties.
As we demonstrate in Sec.~\ref{sec:analysis}, even nodes coming from rather small ego-networks of actors played in different movie genres (here, genre = network class) are surprisingly distinguishable based on a few simple node's properties such as degree, centrality and coreness. 
The fact that nodes can be ``traced back'' to their network class is notable on its own, but in Sec.~\ref{sec:applications} we also consider two applications: (i) whole-network classification from small samples of nodes, and (ii) network bootstrapping (growing networks from small seeds) by iteratively attaching and pruning nodes.

Finally, we note that although there exists a large body of literature on node embedding methods -- which, similar to network embedding, seeks to automatically embed nodes -- these are mainly focused on preserving a node's neighborhood or role in the graph~\cite{cao2015grarep,grover2016node2vec,perozzi2014deepwalk, tang2015line,wang2016structural}.
The goal of node embedding is generally to predict the label of unlabeled nodes in a given graph, assuming that nodes with similar neighborhoods or roles in the network will be labeled the same. 
Here, however, we seek to determine whether nodes from \textit{different} networks can be distinguishable based on their structural features. Specifically, we cannot compare the neighborhoods of these nodes since they come from completely different networks. 

\section{Methodology} \label{sec:methods}
To characterize the structural properties of a given node $u$ in a sparse,  connected network with $N$ nodes and $M$ edges, we use the following network characteristics:
\begin{itemize}
\item {\bf degree centrality}, the fraction of edges adjacent to $u$ (the degree of $u$ divided by the number of nodes, $N$). Computational complexity: $\mathcal{O}(1)$ for each node. 

\item {\bf clustering coefficient}, the density of edges among $u$'s neighbors, i.e., the number of $u$'s friends that are friends among themselves divided by the total number of friend pairs. Computational complexity: $\mathcal{O}({k_u}^2)$ per node of degree $k_u$, or $\mathcal{O}(1)$ assuming the network is sparse and the maximum node degree is independent of $N$.

\item {\bf betweenness centrality}, the flow through a node assuming each pair of nodes $s,t$ exchanges one unit of flow equally split between all shortest paths connecting $s$ and $t$. Thus, betweenness centrality measures the extent to which a node `sits' on shortest paths in the network. Computational complexity: $\mathcal{O}(NM)$ for the entire network~\cite{brandes2001faster}. Betweenness centrality can also be estimated by searching the graph from a small sample of $r$ nodes, yielding $\mathcal{O}(rM)$.
In either case, the computation can be done in parallel (graph search in parallel from each node).

\item {\bf eigenvector centrality}, recursively identifying a node's centrality in terms of the the sum of $u$'s neighbors' centralities. Eigenvector centrality corresponds to the components of the principal eigenvector of the adjacency matrix. Computational complexity: $\mathcal{O}(N+M)$.

\item {\bf closeness centrality}, the reciprocal of average shortest path length from $u$ to nodes. Computational complexity: $\mathcal{O}(M)$ for each node, $\mathcal{O}(NM)$ for the entire network. Alternatively, closeness centrality can be approximated by searching the network from a few `pivot' nodes and approximating other distances using those pivots.

\item {\bf coreness}, the shell index of $u$ is defined as the maximum $k$ such that $u$ belongs to the k-core of the network, which is obtained by recursively deleting nodes with degree less than $k$. Thus, $u$ belongs to the $k$-core if it has at least $k$ neighbors, each with at least $k$ neighbors and so on. Computational complexity: $\mathcal{O}(M)$ for the entire network.

\item {\bf link diversity}, the fraction of $u$'s neighbors outside its own community. Communities are detected using the Leiden algorithm~\cite{traag2019louvain} at default resolution. Computational complexity: $\mathcal{O}(N+M)$ on average. 

\end{itemize}

Once the network features are selected, we use a random forest (RF) classifier, which accounts for correlated features thus allowing for more flexibility with feature space partitioning~\cite{ho1995random}. The network characteristics mentioned above are general and were chosen to cover both local (e.g.\ degree centrality, clustering coefficient) and global (e.g.\ betweenness centrality, link diversity) structural information. 
The method itself is domain-agnostic, but one could consider more or other domain-specific features including nodal attributes (e.g.\ the sex or age of a person in a social network) or dynamic state of nodes/edges (e.g.\ infected or not by an epidemic/rumor).
In each of the tables reporting accuracy values (percentage of correctly classified nodes) in Sec.~\ref{sec:analysis}, we present results for the same experiment but with a \textit{lightweight classifier} that uses only linear-time complexity network features (i.e., removing betweenness and closeness).

\section{Data} \label{sec:data}

To test classifier performance, we used various datasets of empirical networks and random network models.\footnote{The code to download data , parse it, and run each of the five experiments described in Sections~\ref{sec:analysis} and~\ref{sec:applications} is available at  \url{sites.google.com/view/sarayshai-research/publications}}
We use empirical networks in two ways: as complete large-scale networks to distinguish nodes between different networks, and as collections of ego networks (a neighborhood of some focal node with the focal node removed, extracted from the large-scale network) to distinguish nodes from different categories of networks. Each set of ego networks is a category (i.e., labeled by the network from which it was extracted).

\begin{table}[hbp]
\caption{Large-Scale Network Statistics}
\label{table:networks_stats}
\resizebox{\columnwidth}{!}{%
\begin{tabular}{@{}|l|l|l|l|l|l|l|l|@{}}
\toprule
     & \# Nodes & \# Edges & Avg. Degree & Density & Transitivity & Diameter \\ \midrule
FB Cornell     &  18621            & 790753            & 84.93                & 0.0045            & 0.135                & 8                 \\ \midrule
FB UNC      &  18158            & 766796            & 84.45                & 0.0046            & 0.115                & 7                 \\ \midrule
FB UPenn       &  14888            & 686485            & 92.21                & 0.0062            & 0.143                & 9                 \\ \midrule
FB Wesleyan    &  3591             & 138034            & 76.87                & 0.021            & 0.194                 & 7                 \\ \midrule
IMDB Action    &  43974            & 1242285           & 56.5                 & 0.0012            & 0.69                 & 13                 \\ \midrule
IMDB Romance   &  112814           & 4001423           & 70.93                & 0.00062            & 0.511               & 15                 \\ \midrule
IMDB SciFi     &  44483            & 1944367           & 87.42                & 0.0019            & 0.95                 & 13                 \\ \midrule
arXiv CondMat  &  21363            & 91286             & 8.54                 & 0.0004            & 0.26                 & 15                 \\ \midrule
arXiv HepPh    &  11204            & 117619            & 21                   & 0.0018            & 0.66                 & 13                 \\ \midrule
arXiv AstroP   &  17903            & 196972            & 22                   & 0.0012            & 0.32                 & 14                 \\ \bottomrule
\end{tabular}%
}
\end{table}

\begin{table*}[htbp]
\centering
\caption{Network vs. Network Node-Level Classification (with lightweight classifier results in blue)}
\label{table:network_vs_network}
\resizebox{\textwidth}{!}{%
\begin{tabular}{@{}|l|l|l|l|l|@{}}
\toprule
\textbf{}        & \textbf{IMDB Action/Romance/Sci-Fi}                        & \textbf{IMDB Action/Romance}                            & \textbf{IMDB Action/Sci-Fi}                          & \textbf{IMDB Romance/Sci-Fi} \\ \midrule
\textbf{2-fold}  & $98.48 \pm 0.06$ \textcolor{blue}{($97.9 \pm 0.05$)}       &  $99.86 \pm 0.02$ \textcolor{blue}{($99.91 \pm 0.02$)}  &  $97.86 \pm 0.07$ \textcolor{blue}{($96.96 \pm 0.14$)} &  $99.91 \pm 0.02$ \textcolor{blue}{($99.91 \pm 0.02$)}   \\ \midrule
\textbf{10-fold} & $98.73 \pm 0.09$ \textcolor{blue}{($98.2 \pm 0.11$)}       &  $99.92 \pm 0.03$ \textcolor{blue}{($99.94 \pm 0.03$)}  &  $98.21 \pm 0.13$ \textcolor{blue}{($97.3 \pm 0.31$)}  &  $99.94 \pm 0.03$ \textcolor{blue}{($99.94 \pm 0.03$)}  \\ \midrule

& \textbf{arXiv CondMat/HepPh/AstroP} & \textbf{arXiv CondMat/HepPh} & \textbf{arXiv CondMat/AstroP} & \textbf{arXiv HepPh/AstroP} \\ \midrule
\textbf{2-fold}  & $97.96 \pm 0.16$  \textcolor{blue}{($95.76 \pm 0.22$)} & $99.89 \pm 0.04$ \textcolor{blue}{($98.33 \pm 0.16$)}  & $99.9 \pm 0.03$ \textcolor{blue}{($96.64 \pm 0.19$)} & $97.04 \pm 0.2$  \textcolor{blue}{($97.78 \pm 0.15$)} \\ \midrule
\textbf{10-fold} & $98.45 \pm 0.24$  \textcolor{blue}{($96.77 \pm 0.31$)} & $99.92 \pm 0.05$ \textcolor{blue}{($98.71 \pm 0.26$)}  & $99.94 \pm 0.04$ \textcolor{blue}{($97.35 \pm 0.31$)} & $97.74 \pm 0.3$  \textcolor{blue}{($98.4 \pm 0.29$)} \\ \midrule

                 & \textbf{FB Wesleyan/UNC}                                & \textbf{FB Cornell/UPenn}                              & \textbf{FB Wesleyan/Wesleyan}     & \textbf{}  \\ \midrule
\textbf{2-fold}  & $99.87 \pm 0.11$ \textcolor{blue}{($99.94 \pm 0.05$)}   & $91.3 \pm 0.3$ \textcolor{blue}{($89.32 \pm 0.22$)}    & $50.0 \pm 0.0$ \textcolor{blue}{($50.0 \pm 0.0$)}  &            \\ \midrule
\textbf{10-fold} & $99.96 \pm 0.08$ \textcolor{blue}{($99.98 \pm 0.05$)}   & $92.5 \pm 0.54$ \textcolor{blue}{($90.47 \pm 0.5$)}    & $50.0 \pm 0.0$ \textcolor{blue}{($50.0 \pm 0.0$)}   &            \\ \bottomrule
\end{tabular}%
}
\end{table*}

\subsection{Facebook100 Dataset}
The Facebook100 dataset contains complete Facebook networks of 100 American colleges and universities from a single-day snapshot in September 2005~\cite{traud2012social}. It is available to download at \url{https://archive.org/details/oxford-2005-facebook-matrix}. When comparing networks (see Table~\ref{table:network_vs_network}), we have used four complete networks: Cornell University, University of North Carolina at Chapel Hill (UNC), University of Pennsylvania (UPenn), and Wesleyan University. We create 4 sets of ego networks, one for each network, by randomly selecting 1000 nodes that have ego networks with at least 100 nodes. When comparing real to model networks (see Fig.~\ref{fig:real_vs_fake}), we used a subset of 43 networks of size 1,500--10,000 nodes.\footnote{American, Amherst, Bowdoin, Brandeis, Brown, Bucknell, Carnegie Mellon, Colgate, Dartmouth, Emory, Hamilton, Howard, Lehigh, Maine, Michigan, Middlebury, MIT, Oberlin, Pepperdine, Princeton, Rice, Rochester, ``Santa", Simmons, Smith, Swarthmore, Trinity, Tufts, Tulane, UC, UChicago, UCSC, USFCA, Vassar, Vermont, Villanova, Wake, WashU, Wellesley, Wesleyan, William \& Mary, and Williams.}

\subsection{IMDB Actor Collaboration Dataset}
The IMDB actor collaboration network is a bipartite network obtained in 2006 connecting actors and the movies they played in. It is available to download at \url{https://www.cise.ufl.edu/research/sparse/matrices/Pajek/IMDB.html}. The network also contains information about the category of each movie, with 20 categories overall. We extracted actor networks connecting two actors if they played in at least one movie together from the Action, Romance, and Sci-Fi categories  (see network statistics reported in Table~\ref{table:networks_stats}).

\subsection{ArXiv Co-Authorship Dataset}
The arXiv co-authorship dataset encodes scientific collaborations between authors of papers submitted to arXiv.org from January 1993 to April 2003~\cite{leskovec2007graph}. It is available to download at \url{https://snap.stanford.edu/data/#canets}. We used large-scale networks from 3 arXiv categories: Astro Physics (AstroP), Condensed Matter (CondMat), and High Energy Physics - Phenomenology (HepPh) (see network statistics in Table~\ref{table:networks_stats}).

\subsection{Benchmarks}
In addition to the datasets described above, we have also used benchmark datasets that were previously compiled and are commonly used to compare network classification algorithms (all datasets are available to download at \url{https://ls11-www.cs.tu-dortmund.de/staff/morris/graphkerneldatasets}). 

Each benchmark dataset is a collection of ego networks extracted from a large-scale network: IMDB is a collection of ego networks extracted from genre-specific movie-collaboration networks and labeled according to the genre (IMDB-BINARY includes Action and Romance, while IMDB-MULTI includes Comedy, Romance, and Sci-Fi); 
COLLAB is a collection of ego networks of different researchers from each field (High Energy Physics, Condensed Matter Physics, and Astro Physics) labeled according to the field of the ego node; 
REDDIT contains interactions between Reddit users in online discussion threads with networks labeled based on the subreddit that it was extracted from. 

\section{Experiments} \label{sec:analysis}
\subsection{Network vs. Network} \label{sec:exp1}

We begin by testing our model's ability to distinguish between nodes belonging to different networks.
In particular, consider a binary classification task where, given a pair of networks $G_1$ and $G_2$, we train a model to predict whether an unseen node belongs to $G_1$ or $G_2$.
We first compute the network properties of all nodes in each network, label nodes according to the network they belong to ($G_1$ or $G_2$), and divide the nodes into training and test groups.
We perform our tests on two pairs of complete Facebook networks, three pairs and a triplet (that is, multi-label classification) of IMDB actor collaboration networks, and three pairs and a triplet of arXiv co-authorship networks (see Table~\ref{table:networks_stats} for details about each of these networks.)

In Table~\ref{table:network_vs_network} we report results from 10 repeated runs of 2-fold and 10-fold cross validation (numbers represent average and standard deviation of percentage of correctly classified nodes over all folds from all repeated runs). In blue parentheses, we also show results with a lightweight classifier that uses only linear-time complexity network features (i.e., removing betweenness and closeness).
All tests have accuracy of at least 91\% (or at least 89\% with the lightweight classifier), including multi-label classification tests. To further validate our results, we trained a classifier to distinguish between nodes from the same network (Facebook network of Wesleyan University), obtaining 50\% accuracy as expected. In other words, given unseen nodes from two identical copies of the Wesleyan Facebook network, the classifier could not determine whether these nodes belonged to the first or the second one. 

\subsection{Networks Class vs. Networks Class} \label{sec:exp2}

The above experiment tells us that a classifier can be trained to distinguish between nodes in specific networks: given a list of structural features of a node, the classifier can predict (with very high accuracy) the specific network from which this node was `taken'. But one has to bear in mind that the training and test nodes are connected (either directly or by being in the same connected component), potentially making the prediction easier. 
Thus, our next experiment tests the ability to distinguish between nodes coming from networks that belong to different \textit{network classes}, by which we mean a set of networks that can be distinguished (e.g.\ using a whole-network classifier) from another set of networks.
This task is more challenging since the classifier is trained with a collection of nodes from multiple networks in the same class.
Then, given a test node, the classifier has to determine whether this node looks more similar to the training nodes in class $A$ or class $B$. 

To create classes of networks, we extract ego networks from each of the 4 Facebook networks used in Section~\ref{sec:exp1}, randomly choosing 1000 ego networks with at least 100 nodes. The class of each ego network is defined by the large network it came from. We then divide the ego networks into folds and compute the network properties of each node in each ego network. We train the classifier on \textit{all} nodes from all training networks and test it on \textit{all} nodes from all test networks.

\begin{figure}[htbp]
	\centering
	\includegraphics[width=\columnwidth]{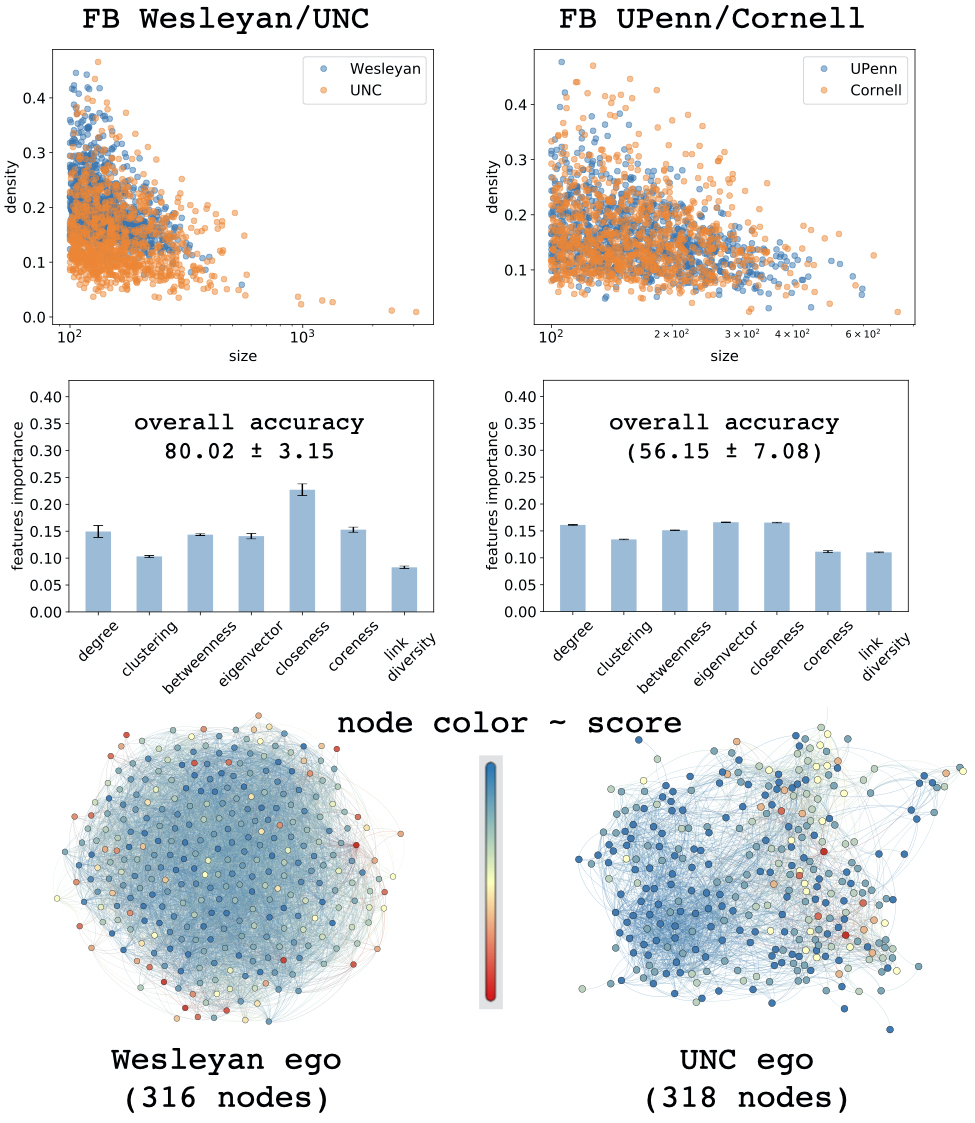}
	\caption{Distinguishing nodes in different network classes. Top row: network density versus network size for each ego network extracted from Wesleyan and UNC whole-networks (see Table~\ref{table:networks_stats}). Middle row: average and standard deviation of feature importances and accuracy values obtained from 10 repeated runs of a 10-fold experiment (100 classifiers overall). Bottom row: network visualization of a Wesleyan ego network (left) and a UNC ego network (right). Both networks were visualized in Gephi~\cite{bastian2009gephi} using the same layout algorithm and same parameters (color spline, etc). Node colors correspond to their classification with correct label being 1 and wrong label being 0 (so red nodes are those that were classified incorrectly).}
	\label{fig:real_vs_real}
\end{figure}

The sizes and densities of the extracted ego networks reported in Fig.~\ref{fig:real_vs_real}~(top row) show similar size/density distributions across different classes used in the same experiment. For example, even though the UNC Facebook network is significantly larger than the Wesleyan Facebook network, the comparable sizes and densities make the classification task nontrivial (i.e. the classifier can't simply infer based on nodes in one class having smaller degrees). 

We perform 10 repeated runs of 10-fold cross validation (i.e., 100 overall) and report average accuracy and standard deviation over all folds and runs (see Fig.~\ref{fig:real_vs_real}, middle row). The accuracy of a given fold in a given run is obtained by comparing the true and predicted label of all nodes from all test networks in that fold. We also show the average weight that the classifier assigns each feature. For both Facebook classifiers, we observe that feature importances are fairly evenly distributed (no feature gets more than 25\% of the weight). However, we observe a significant difference between the Wesleyan/UNC classifier's performance and the UPenn/Cornell classifier even though their ego networks span similar size/density. (We note that the Wesleyan/UNC comparison is between a small, private liberal arts school and a flagship public university, while the UPenn/Cornell comparison is between two Ivy League schools of similarly-sized undergraduate populations, which may be part of the reason it appears to be a harder task.)
In Fig.~\ref{fig:real_vs_real}~(bottom row) we show a network visualization of two ego networks -- one from Wesleyan (left) and one from UNC (right) -- where nodes are colored according to their classification score: blue nodes are classified ``correctly'' (their score is similar to their actual label) and red nodes are classified ``incorrectly'' (their score is similar to the wrong label).  
Both visualizations were performed with the same layout algorithm and the exact same parameters. Although it is hard to see specific network properties from such a visualization, it is notable that misclassified nodes tend to be on the periphery of the network or between denser modules (or communities).
Another thing to note in this visualizarion is that the UNC ego networks generally tend to organize into three or more communities while the Wesleyan ego networks are much more homogeneous. This is illustrated in the two networks shown in the figure, but was also observed in general (e.g.\ the average modularity~\cite{newman2006modularity} of the UNC ego networks is 0.35 compared with 0.28 for Wesleyan.)

\begin{table}[h]
\centering
\caption{Node-Level Classification in Benchmark Datasets (with lightweight classifier results in blue)}
\label{table:benchmarks}
\begin{tabular}{@{}|l|l|l|@{}}
\toprule
COLLAB           & IMDB-BINARY      & IMDB-MULTI       \\ \midrule
\shortstack{$79.83  \pm 1.95$   \\ \textcolor{blue}{$(80.12 \pm 1.99)$}} & \shortstack{$72.33 \pm 3.65$  \\ \textcolor{blue}{$(72.77 \pm 3.6)$}}  & \shortstack{$52.03 \pm 3.62$   \\ \textcolor{blue}{($52.22 \pm 3.6$)}}  \\ \midrule
REDDIT-BINARY    & REDDIT-MULTI-5K  & REDDIT-MULTI-12K \\ \midrule
\shortstack{$82.8 \pm 2.62$ \\ \textcolor{blue}{$(76.41 \pm 3.11)$}}  & \shortstack{$ 41.21 \pm 1.95$ \\ \textcolor{blue}{$(36.56 \pm 1.68)$}} & \shortstack{$32.89  \pm 2.12$ \\ \textcolor{blue}{$(24.79 \pm 1.99)$}} \\ \bottomrule
\end{tabular}%
\end{table}

\begin{table*}[hbtp]
\centering
\caption{Feature Importances of Node-Level Classification in Benchmark Datasets}
\label{table:benchmarks_features}
\begin{tabular}{@{}|l|l|l|l|l|l|l|@{}}
\toprule
          & COLLAB                      & IMDB-BINARY                  & IMDB-MULTI                & REDDIT-BINARY              & REDDIT-MULTI-5K           & REDDIT-MULTI-12K   \\ \midrule
Degree           & $13.11 \pm 1.45$            & $16.01 \pm 0.62$             & $19.42 \pm 0.83$          & $\mathbf{42.36 \pm 1.25}$  & $\mathbf{34.26 \pm 0.23}$ & $\mathbf{32.54 \pm 0.27}$ \\ \midrule
Clustering       & $5.12 \pm 0.31$             & $3.63 \pm 0.24$              & $3.52 \pm 0.14$           & $0.87 \pm 0.03$            & $0.78 \pm 0.02$           & $0.98 \pm 0.02$   \\ \midrule
Betweenness      & $5.35 \pm 0.22$             & $4.22 \pm 0.29$              & $3.87 \pm 0.17$           & $7.29 \pm 0.24$            & $7.0 \pm 0.08$            & $7.12 \pm 0.13$   \\ \midrule
Eigenvector      & $23.28 \pm 1.57$            & $\mathbf{33.82 \pm 0.65}$    & $\mathbf{26.63 \pm 0.52}$ & $21.66 \pm 0.65$           & $27.32 \pm 0.2$           & $28.36 \pm 0.23$  \\ \midrule
Closeness        & $11.43 \pm 0.83$            & $12.06 \pm 0.42$             & $12.44 \pm 0.42$          & $24.89 \pm 0.83$           & $27.98 \pm 0.18$          & $28.55 \pm 0.19$  \\ \midrule
Coreness         & $\mathbf{34.36 \pm 1.21}$   & $23.39 \pm 0.95$             & $25.86 \pm 0.99$          & $1.25 \pm 0.1$             & $1.27 \pm 0.09$           & $0.77 \pm 0.05$   \\ \midrule
Link diversity   & $7.35 \pm 0.37$             & $6.88 \pm 0.32$              & $8.26 \pm 0.28$           & $1.68 \pm 0.07$            & $1.39 \pm 0.04$           & $1.67 \pm 0.03$   \\ \bottomrule
\end{tabular}%
\end{table*}
Finally, we perform the same experiment on benchmark datasets commonly used for whole-network classification. We test our model's ability to tell apart nodes coming from networks in different classes:
Table~\ref{table:benchmarks}, reports results for the regular and lightweight classifier, and  Table~\ref{table:benchmarks_features} reports feature importances for each dataset, observing that COLLAB relies most heavily on coreness while REDDIT and IMDB rely most heavily on degree and eigenvector centrality respectively.
To compare our results with the state-of-the-art on these benchmarks, in Section~\ref{sec:applications} we use our node-level classifier to classify whole networks by taking the average score of a sample of the nodes, which yields comparable accuracy values. In contrast, we are unaware of any node-by-node results in the literature to compare to our results in Table~\ref{table:benchmarks}.

\subsection{Real vs. Model Networks} \label{sec:exp3}
We have established above the ability of a node-level classifier to distinguish between specific networks and between classes of networks. Next, we test its ability to differentiate real from model data. Specifically, for each network in the Facebook100 dataset (limiting this experiment to networks with 1,500--10,000 nodes), we create five model networks: (i) a random Erd\H{o}s-R\'enyi (ER) network~\cite{erdHos1960evolution}
with the same number of nodes and edges, (ii) a configuration model network~\cite{bollobas1980probabilistic,fosdick2018configuring} with the same degree sequence, (iii) a scale-free Barab\'{a}si-Albert (BA) network~\cite{barabasi1999emergence} with the same number of nodes and edges, 
(iv) a small-world Watts–Strogatz (WS) network~\cite{watts1998collective} with the same number of nodes and edges with rewiring probability $p=0.1$, and (v) a BA network with clustering according to the Holme-Kim model~\cite{holme2002growing}. 

\begin{figure*}[htbp]
	\centering
	\includegraphics[width=.95\textwidth]{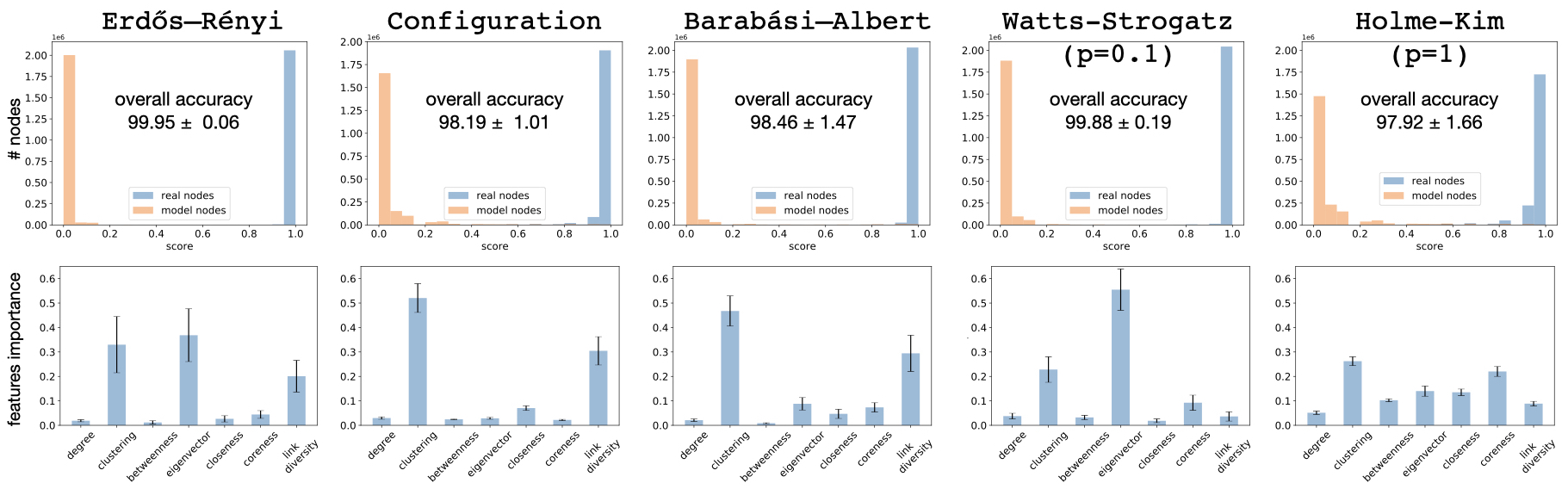}
	\caption{Distinguishing real from model nodes. Distributions of scores (and overall accuracy values) for real and model nodes for each of the five random network models described in the text (top row) and  feature importances for the corresponding classifiers (second row). Results shown are obtained by averaging 100 classifiers (10  independent 10-fold cross-validation). The distributions show all test nodes from all 10 runs.}
	\label{fig:real_vs_fake}
\end{figure*}

\begin{figure}[htb]
	\centering
	\includegraphics[width=.5\textwidth]{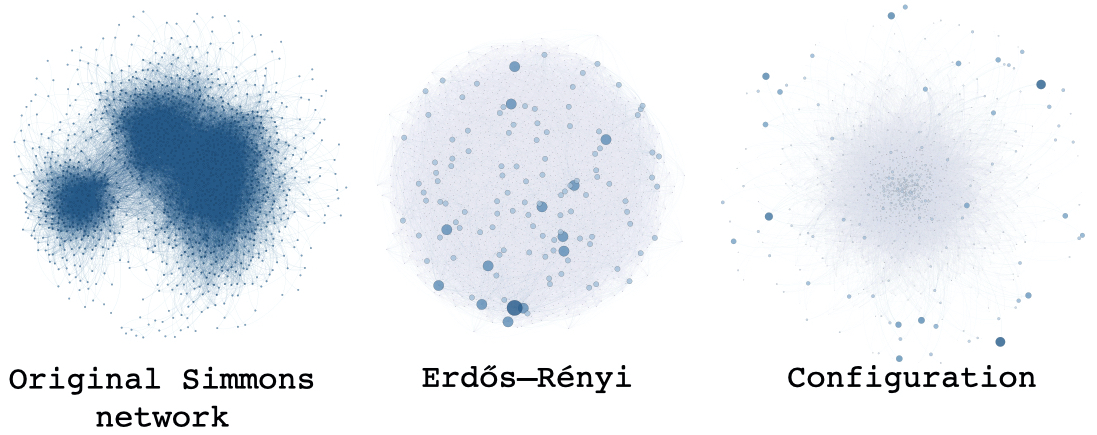}
	\caption{Network visualizations. Original Simmons network (left), a corresponding ER network with the same number of nodes and edges (middle), and a corresponding configuration model with the same degree sequence (right). All networks were visualized in Gephi~\cite{bastian2009gephi} using the same layout algorithm.
	Nodes in the model networks are colored and sized according to their score (darker large nodes look more `real' to the classifier).}
	\label{fig:real_real_vis}
\end{figure}

We perform 10 repeated runs of 10-fold cross validation, where each run creates new random networks (from each of the five models) corresponding to each real Facebook network (see Fig.~\ref{fig:real_vs_fake}). First, we note that none of the five models were particularly successful at ``fooling'' the classifier into thinking that real nodes are from random models or that random model nodes are real (overall accuracy is above 97.9\% in all models). This is another potential application of a node-level classifier: to evaluate network models by how `real' the nodes in the network look. In this regard, the Holme-Kim model (with probability $p=1$ of closing a triangle after every random edge addition) leads to the lowest accuracy: the classifier occasionally mistakes model nodes as real and vice versa.

Second, we observe the configuration and BA classifiers behave similarly in giving clustering and link diversity highest importance with very little to degree/eigenvector centrality, while the ER and WS classifiers give eigenvector centrality and clustering coefficient the most importance. We also note that Holme-Kim's preferential attachment model with triangle closure yields the most evenly-distributed feature importances.

Finally, in Fig.~\ref{fig:real_real_vis} we visualize a real network (we chose one of the smallest networks in the dataset, the Simmons University network) and corresponding example ER and configuration networks with nodes colored and sized according to their score for looking `real' (large, dark nodes scored high). We observe that high-score nodes in the random ER network tend to be distributed all over the network while high-score nodes in the configuration model network tend to be in the periphery with few slightly larger nodes in the network's core.

\section{Applications} \label{sec:applications}
\subsection{Whole Network Classification}
As mentioned in Section~\ref{sec:exp2}, our node-level classifier can be used to classify whole networks by defining a network score as, for example, the average score of its nodes, a weighted average (e.g.\ giving high-centrality nodes more weight), or the fraction of nodes with score above some threshold. 

\begin{figure}[hbt]
	\centering
	\includegraphics[width=0.98\columnwidth]{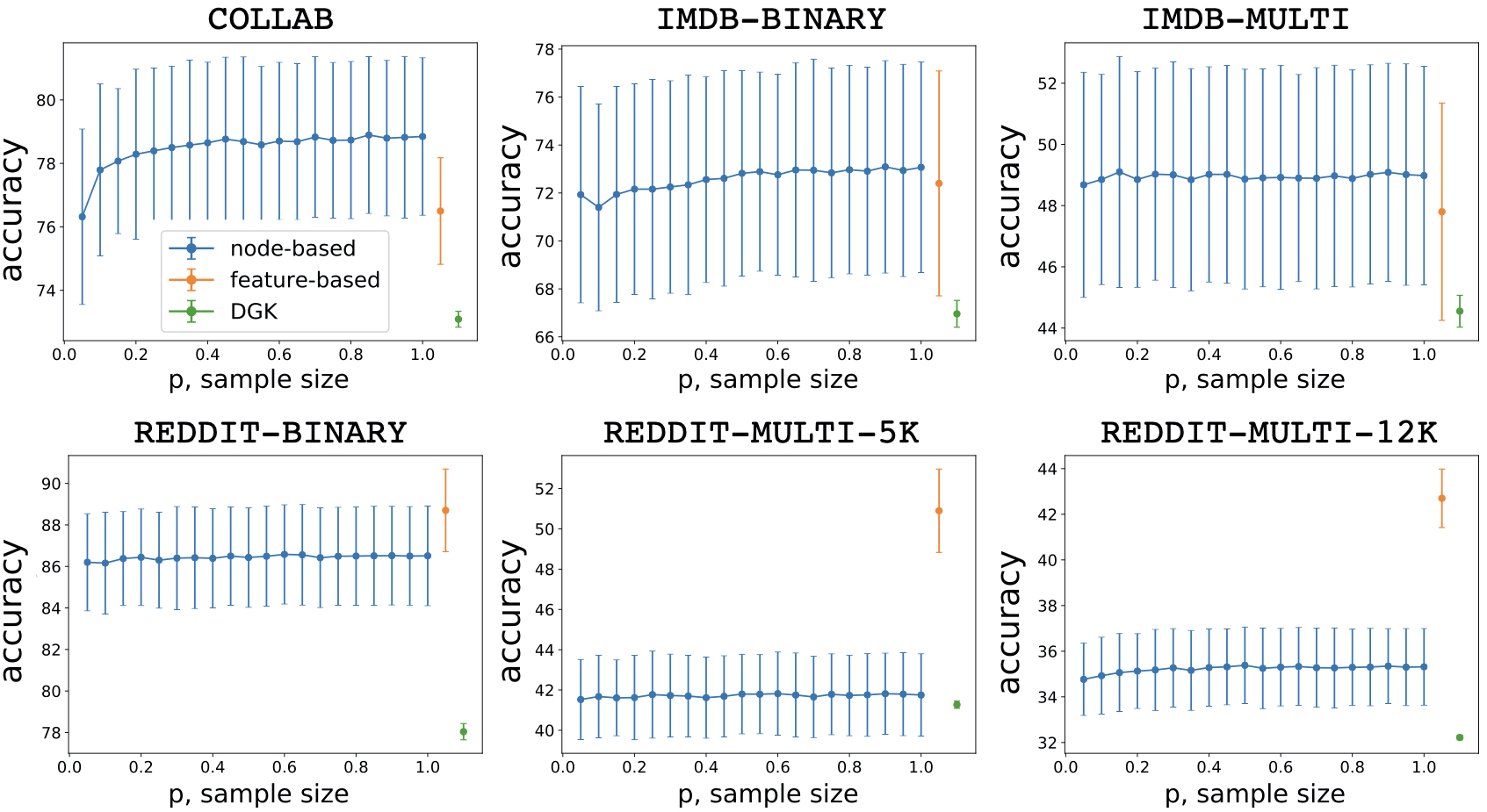}
	\caption{Whole-network classification. For each of the benchmark datasets discussed in Section~\ref{sec:exp2}, we compare whole-network classification accuracy of feature-based classification~\cite{barnett2019endnote} and Deep Graph Kernels (DGK)~\cite{yanardag2015deep} with our node-level classifier trained on single nodes. For the node-level classifier, we report average accuracy and standard deviation from 10 independent runs of a 10-fold cross validation where network scores are obtained by averaging the score of a random sample of a fraction $p$ of their nodes.}
	\label{fig:app1}
\end{figure}

In Fig.~\ref{fig:app1} we compare the performance of network classification by averaging over samples of our node-level classifier versus Deep Graph Kernels (DGK)~\cite{yanardag2015deep} and feature-based classification~\cite{barnett2019endnote}. That is, we define the score of a given network as the average score of a random sample of its nodes, denoting by $p$ the fraction of nodes in the sample. For $p=1$, the score of a network is just the average score of all its nodes.

Note that the classifier is still trained exactly as before (that is, the training data is from single-node features, see Table~\ref{table:benchmarks_features}) but the accuracy is obtained differently. Instead of comparing each node's label with its prediction, we now compare a whole network's label with the prediction from the average score of its nodes (as a function of the sample size $p$). Thus, given a fraction $p$ of the nodes to sample, we perform our regular 10-fold cross validation but now the score of a given test network on each label is the average score of a random sample (of fraction $p$) of its nodes. The predicted label of the entire network is the label on which the average score is the highest. As before, we perform 10 independent runs of 10-fold cross validation. Feature-based and DGK accuracy is obtained in the same way (by comparing whole-network classification with its label) and does not depend on $p$.

We observe that all six benchmark datasets behave similarly: the accuracy of whole-network classification converges quickly with increasing sample size to values similar to those reported in Table~\ref{table:benchmarks} for node-level classification. For example, the COLLAB benchmark starts with  around 76.3\% and quickly increases above $78.5$\% when sampling 40\% of the nodes (Fig.~\ref{fig:app1} top left). The corresponding node-level accuracy in Table~\ref{table:benchmarks} is $79.83 \pm 1.95$. In the REDDIT-BINARY benchmark data, the accuracy is very stable already at small sample sizes and eventually converges to 86.51\% for $p=1$ (Fig.~\ref{fig:app1} bottom left) while the corresponding accuracy in Table~\ref{table:benchmarks} is 82.8\%. 

We note our results are comparable with both feature-based classification and DGK, with  average accuracy from the node-level classifier consistently outperforming DGK in all datasets. The feature-based classifier is performing better on the multi-label classification datasets (REDDIT-5K and REDDIT-12K). This might be explained by noting that the feature-based classifier uses properties of the whole benchmark ego network at a time whereas in our approach we are averaging node-level features. Taken together, this experiment demonstrates that a node-level classifier that is trained on nodes from different network classes can also be used to classify whole test networks by averaging the score of some of its nodes.

\subsection{Network Bootstrapping}\label{sec:net_growth}
Finally, we show how to use a node-level classifier to grow ``real-looking networks'' from an original network (sometimes called \textit{graph generation}~\cite{you2018graphrnn}). We start by training a classifier to distinguish configuration model networks from a given real network as in previous sections. We then grow by iteratively adding and pruning nodes with score less than some threshold (i.e.\ nodes that don't look ``real enough'' to the classifier).

Specifically, we consider two node attachment mechanisms to add new nodes to the network: (i) Vertex Copy: when a node is added, we randomly choose an existing node and copy fraction $\beta$ of its $k$ links. The remaining $k(1-\beta)$ links are randomly selected. (ii) Triadic Closure: when a new node is added, it obtains a degree $k$ by sampling a degree from the original network's degree sequence and multiplying it by $n/N$ (ratio between grown and original network). For each of the $k$ links, with probability $1-\beta$ we are choosing a random neighbor, and with probability $\beta$ we are choosing a neighbor of an existing neighbor to close a triangle). 

Let $G$ denote the original network, and $g$ the grown network. To add a new node $u$ to $g$ according to the Vertex Copy attachment mechanism, we randomly select an existing node $v$ in $g$ and create a node $u$ with the same degree, $k=deg_g(v)$. Then, we randomly select fraction $\beta$ of $v$'s links and copy them. The remaining neighbors of $u$ are chosen at random:

\begin{algorithm}
\caption{VertexCopy(g, beta, MAXTRIALS)}
\begin{algorithmic}[1]
\STATE v = RandomNode(g)
\STATE k = Deg(g, v)
\STATE count = 0
\STATE FRIENDS = Set()
\WHILE{len(FRIENDS) $<$ k and count $<$ MAXTRIALS}
\STATE r = Uniform(0,1)
\IF{r $<$ beta}
\STATE w = RandomNode(Neighbors(g, u))
\ELSE
\STATE w = RandomNode(g)
\ENDIF
\STATE FRIENDS.add(w)
\STATE count = count + 1
\ENDWHILE
\end{algorithmic}
\end{algorithm}

MAXTRIALS is the maximum number of times that we will try to find friends for the new node $u$. In some very rare cases, we might fail to find enough friends while keeping the graph simple (for example, if $v$ has very few friends that were already selected as the `random neighbors' in line 10. 

To add a new node $u$ to $g$ according to the Triadic Closure attachment mechanism, we first define the degree of $u$ to be $k=k_i \frac{n}{N}$ where $k_i$ is a randomly selected degree from $G$'s degree sequence, $n$ is the number of nodes in $g$ and $N$ is the number of nodes in $G$. Then, we select $k$ neighbors either at random (with probability 1-$\beta$) or in such a way that closes a triangle (with probability $\beta$):

\begin{algorithm}
\caption{TriadicClosure(g, beta, K[1....N], MAXTRIALS)}
\begin{algorithmic}[1]
\STATE k = RandomChoice(K) * NumNodes(g) / N
\STATE count = 0
\STATE FRIENDS = Set()
\WHILE{len(FRIENDS) $<$ k and count $<$ MAXTRIALS}
\STATE r = Uniform(0,1)
\IF{r $<$ beta and len(FRIENDS) $>$ 0}
\STATE v = RandomChoice(FRIENDS)
\STATE w = RandomNode(Neighbors(g, v))
\ELSE
\STATE w = RandomNode(g)
\ENDIF
\STATE FRIENDS.add(w)
\STATE count = count + 1
\ENDWHILE
\end{algorithmic}
\end{algorithm}

\begin{figure}[htbp]
	\centering
	\includegraphics[width=.92\columnwidth]{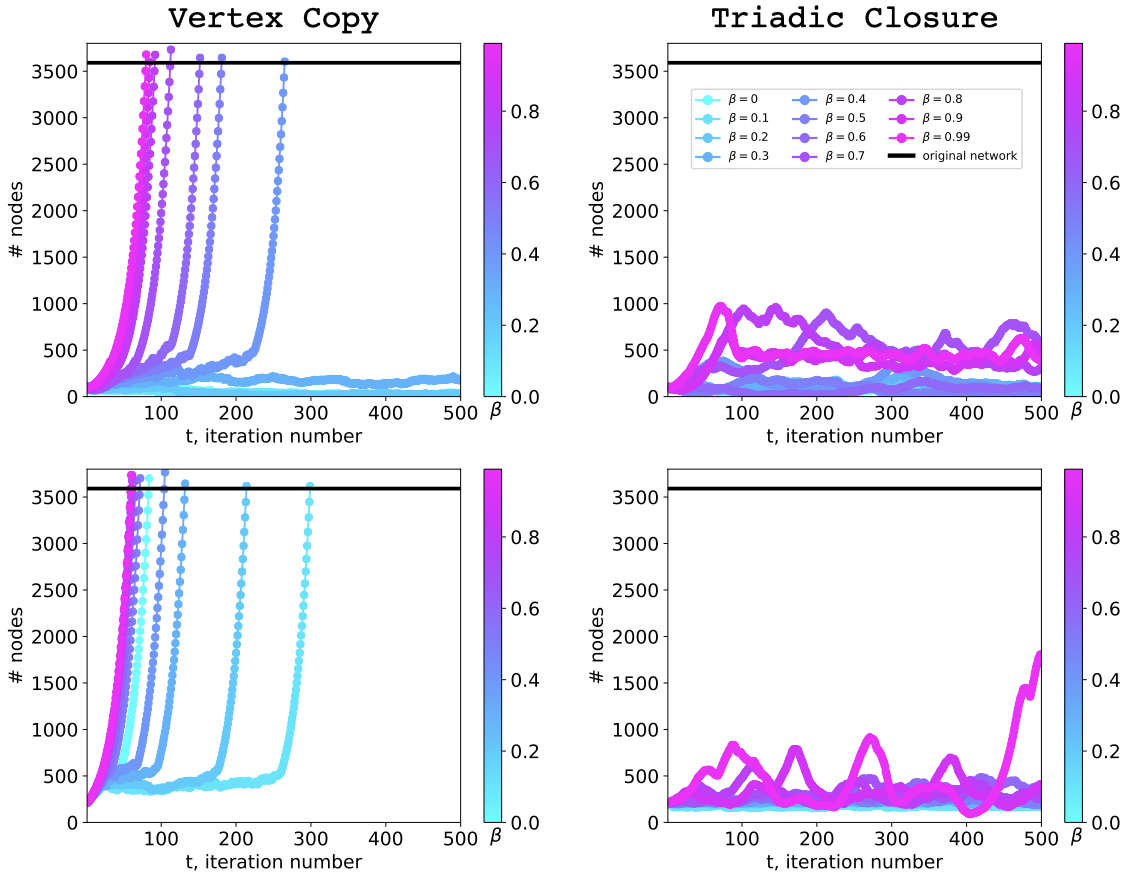}
	\caption{Network Growth. Number of nodes in the grown network, $\mathbf{n}$, as a function of time. Each iteration, $\mathbf{n \cdot 0.05}$ nodes are added according to the Vertex Copy (left) and Triadic Closure (right) mechanism, and then all nodes are re-classfied and removed if their score is less than 0.8. Each row corresponds to one seed network that was used for all $\mathbf{\beta}$ values for each attachment mechanism. The black line shows the number of nodes in the original Wesleyan Facebook network.}
	\label{fig:network_growth}
\end{figure}

In Fig.~\ref{fig:network_growth}, we show the number of nodes in the grown network versus time for each attachment mechanism. The original network is the Wesleyan Facebook network (N=3591 nodes). The initial seed network is an ego network (without the ego node) chosen at random from those containing at least 100 nodes. Each iteration, we grow the network by 5\%, adding $n \cdot 0.05 $ new nodes and re-calculate scores of all the nodes. Nodes that score at least $0.8$ (that is, the Wesleyan vs.\ Configuration classifier assigns these nodes at least 80\% chance of being 'Wesleyan nodes') are kept in the network and all other nodes are removed. We run the experiment for 500 iterations or until the grown network has reached $N$ nodes.

Fig.~\ref{fig:network_growth} shows results from two representative runs of this experiment. Each run has the same seed network for different values of $\beta$; plots in the same row have the same seed network but different attachment mechanism. The number of nodes per iteration highly depends on the seed network, but the trajectory is qualitatively similar across runs. We see that the Vertex Copy mechanism is able to quickly grow networks as large as the original network (especially for high values of $\beta$) starting from a small network of only about 100 nodes. Thus, the resulting networks can potentially be used for statistical tests by measuring the probability of observing a `real-looking' network with a certain characteristics (e.g.\ a certain modularity score). However, the Triadic Closure mechanism typically hits an upper limit, even for large $\beta$ values, meaning that it can no longer add nodes that look like `Wesleyan nodes.'

\section{Conclusions and future work}
We studied whether a classifier can be trained to predict the network category of single nodes based on their structural characteristics.
We tested our framework in various settings to distinguish nodes from different large-scale networks, different classes of ego networks, and classes of empirical and random networks. We also considered potential applications of our framework, including network classification and network bootstrapping. Using benchmark datasets commonly used for network classification, we demonstrated that our framework accurately predicts a network's category by averaging the scores of (some small sample of) individual nodes (Fig.~\ref{fig:app1}).

Another application worth investigating is using our framework to create new random network models by rewiring edges and rejecting changes that significantly decrease the overall classification score of the network. 
The fact that nodes from different network classes are distinguishable based on a few simple network measures is not trivial, pointing to interesting future directions to study, including the rate at which nodes lose their ``realness'' under random rewiring, and how the number of randomly-selected nodes required for accurate whole-network classification scales with network size.

\bibliographystyle{IEEEtran}

\end{document}